\documentclass[journal = aesccq,manuscript=article]{achemso}
\setkeys{acs}{articletitle = false}
\usepackage{natbib}
\newcommand{\kms}{\mbox{km~s$^{-1}$~}}

\author{Nicolas Biver}
\email{nicolas.biver@obspm.fr}
\affiliation[LESIA]
{LESIA, Observatoire de Paris, PSL Research University, CNRS, 
          Sorbonne Universit\'e, Univ. de Paris, 
          Sorbonne Paris Cit\'e, 5 place Jules Janssen, F-92195 Meudon, France}
\phone{+33 1 45 07 78 09}
\author{Dominique Bockel\'ee-Morvan}
\email{dominique.bockelee@obspm.fr}
\affiliation[LESIA]
{LESIA, Observatoire de Paris, PSL Research University, CNRS, 
          Sorbonne Universit\'e, Univ. de Paris, 
          Sorbonne Paris Cit\'e, 5 place Jules Janssen, F-92195 Meudon, France}

\title[COMs in comets]{Complex organic molecules in comets from remote-sensing observations at millimeter wavelengths}
\abbreviations{COM,ISM,OCC,JFC}

\begin{document}

\begin{abstract}
  Remote observations of comets, especially using high spectral
  resolution millimeter spectroscopy, have enabled the detection of over
  25 molecules in comets for the last twenty years. Among the molecules
  identified at radio wavelengths, complex organic molecules (COMs) such as
  acetaldehyde, ethylene-glycol, formamide, methyl-formate or ethanol have been
  observed in several comets and their abundances relative to water and methanol
  precisely determined. Significant upper limits on the abundance of several
  other COMs have been determined and put constraints on the dominant isomer
  for three of them. The abundances measured in comets are generally of
  comparable order of magnitude as those measured in star-forming regions,
  suggesting that comets contain preserved material from the presolar cloud
  from which the solar system was born.
\end{abstract}

\section{Introduction}
\begin{figure}
\centering
\resizebox{14cm}{!}{\includegraphics[angle=180]{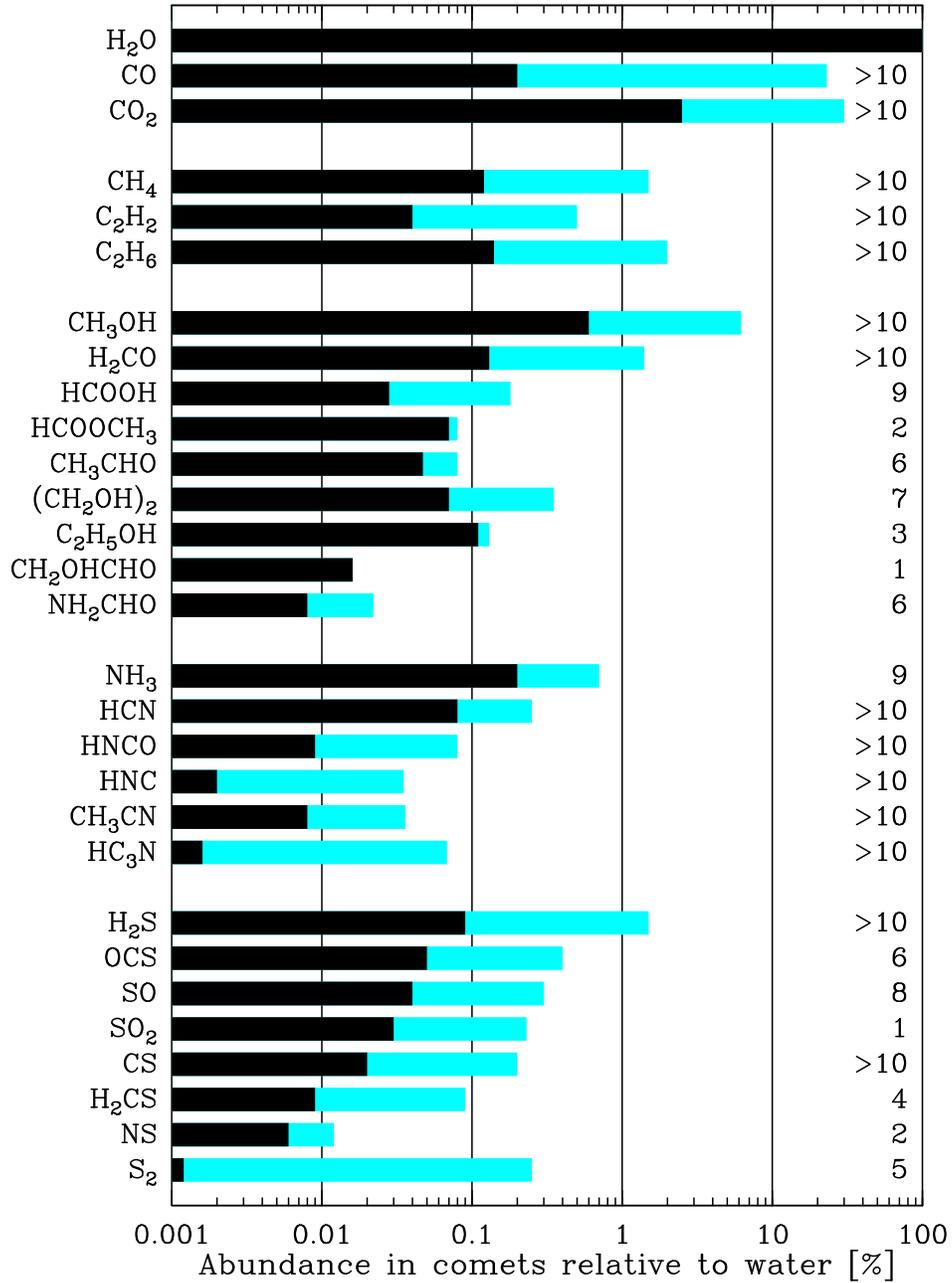}}
\caption{Relative abundance of molecules detected in comets remotely -
updated from \cite{Boc17}, with data on comets recently observed.}
\label{fighistoall}
\end{figure}

The presence of numerous complex organic molecules 
(COMs; defined as those containing 6 or more atoms)
around protostars shows that star formation is accompanied by an
increase of molecular complexity. These COMs may be part of the
material from which planetesimals and ultimately planets formed
\cite{Her09}.
Comets are the most pristine remnants of the formation of the
Solar System. They sample some of the oldest and most
primitive material in the solar system, including ices, and are
thus our best window into the volatile composition of the solar
proto-planetary disk. Comets may also have played a role in the
delivery of water and organic material to the early Earth.
Their composition has been studied both remotely and in-situ using
mass spectroscopy aboard spacecraft such as the Rosetta Orbiter
Spectrometer for Ion and Neutral Analysis (ROSINA) which identified
over 60 molecules in the coma of comet 67P/Churyumov-Gerasimenko
in 2014-2016 \cite{Alt17,Ler15,Bie15,Rub15,Alt16,Cal16,Rub18}.
The recent years have also seen significant improvement in the sensitivity
and spectral coverage of infrared to millimetre telescopes. Ground-based
and remote observations have enable the detection of over 25 molecular
species in many comets and showed that their relative abundances are diverse
\cite{Del16,Biv17,Boc17}. Figure~\ref{fighistoall} provides an overview
of the molecules detected in comets by remote techniques and the range
of measured abundances relative to water.

In this paper we include data on both recently observed comets
  and previously published papers\cite{Biv14,Biv15,Biv16,Biv17,Boc17}
and further analysis
of the comet C/2014~Q2 (Lovejoy) dataset\cite{Biv15} on upper limits on
complex molecules. Some of the previous results obtained for comet
C/1995 O1 (Hale-Bopp) \cite{Cro04} have also been updated with up-to-date
models and molecular parameters\cite{CDMS,JPL}.
We also include some of the results of the observations
of comet 46P/Wirtanen conducted with the IRAM-30m telescope in december 2018
\cite{Biv19}. These represent the most sensitive remote molecular survey of a
Jupiter Family comet to date.

\section{Millimeter observation of comets}
Excepted for the molecules that lack a permament dipole moment, having
significant (vibrational) transitions only in the infrared, the most efficient
technique of detecting remotely complex molecules is via radio spectroscopy of
their rotational lines.
Provided that the molecular spectra are well known, the millimeter to
submillimeter technique can unambiguously distinguish molecules and their
isomers thanks to its high spectral resolution capability
($\Delta\nu/\nu \leq 10^{-6}$)
and owing to the narrow cometary lines ($\approx$2\kms). The very high spectral
resolution of the radio technique also enables to resolve the velocity
profiles of the lines and measure outflow speeds.
Today about 200 molecules have been detected in the interstellar medium
or circumstellar shells, for which laboratory spectra have been established
with sufficient accuracy for unambiguous spectral identification
\cite{CDMS,JPL}.

\section{Complex organic molecules detected in comets}
Methanol was first identified in the ISM in sources Sgr A and B2 in 1970
\cite{Bal70}.
It was the first COM (as defined) identified in comets and has now
been detected both in the infrared and in the radio in over 40 comets
\cite{Del16,Biv17}. Since the apparitions of the bright comets
C/1996~B2 (Hyakutake) and C/1995~O1 (Hale-Bopp) which resulted in the first
identification of many COMs in comets, several have
been identified in comets over the past twenty years
\cite{Boc00,Cro04,Cro04b,Biv14,Biv15}. The abundance
of a molecule in cometary comae tends to decrease \cite{Cro04} with increasing
molecular complexity.
In addition, the rotational energy of the molecule is spread over an
increasing number of levels when the number of atoms increases and the
individual rotational lines tend to be fainter and fainter. However, thanks
to the improved sensitivity and wavelength coverage of the millimeter receivers
and spectrometers the
detection of COMs in comets less active than, e.g. Hale-Bopp, is now possible
\cite{Biv14,Biv15}. Table~\ref{tabcoms} provides an overview of the
abundances of the complex organic molecules detected to date in comets by
remote sensing observations.

\begin{figure}
\centering
\resizebox{13cm}{!}{\includegraphics[angle=0]{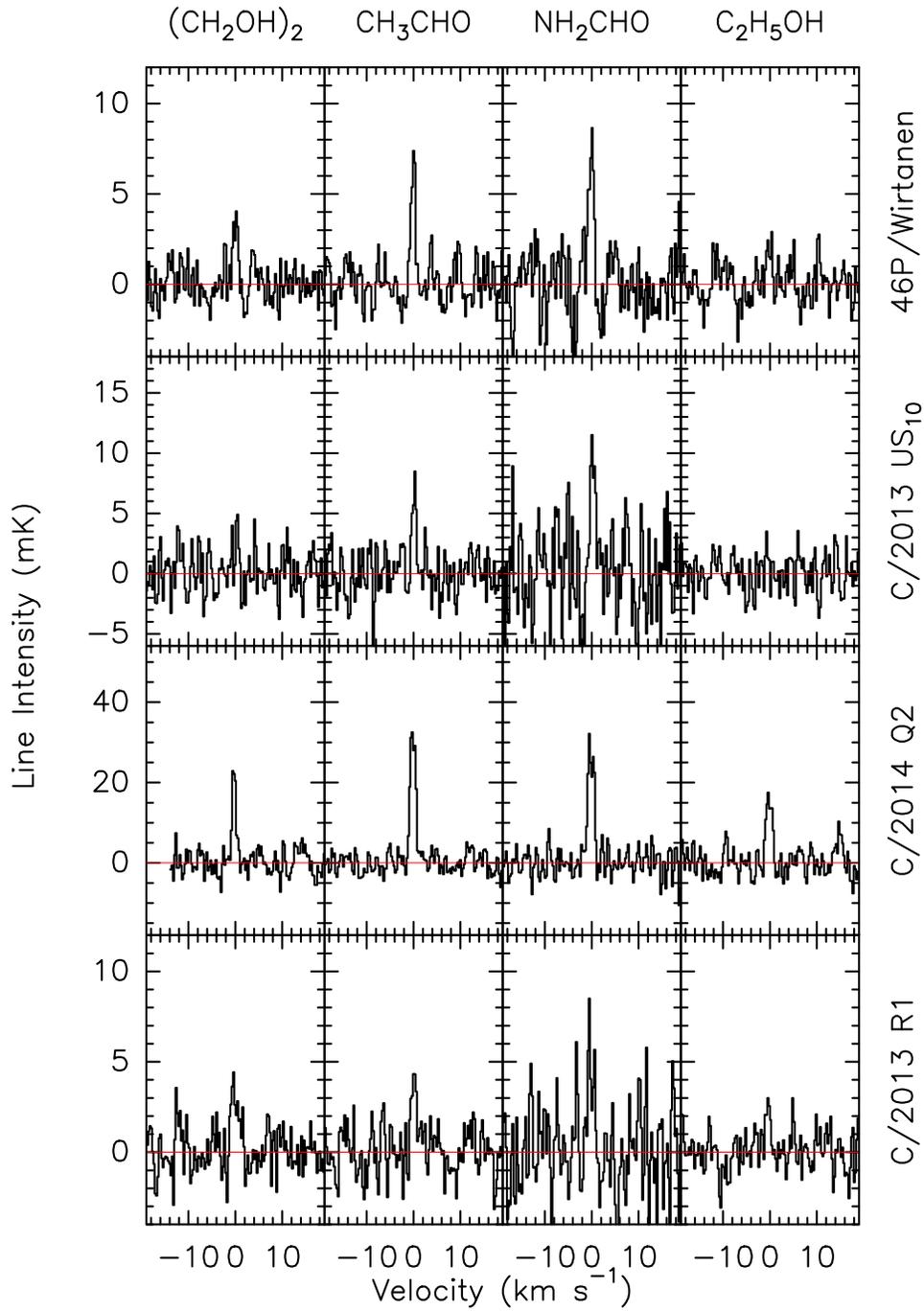}}
\caption{Sample spectra of COMs recently observed in comets. Spectra
  of comet C/2014~Q2 are from \cite{Biv15}. All spectra are the average of
  several millimeter lines expected to be of similar intensity \cite{Biv14}.
  The lines were observed with the IRAM-30m radio telescope between 210 and 272 GHz.
  For comets C/2013~R1, C/2013~US$_{10}$ and 46P, the (CH$_2$OH)$_2$
  and C$_2$H$_5$OH lines are the average of $\approx80$ transitions,
  $\approx65$ transitions for CH$_3$CHO and
  $\approx18$ transitions for NH$_2$CHO.}
\label{figspeccoms}
\end{figure}

Abundances for comet C/1995 O1 (Hale-Bopp) in Tables~\ref{tabcoms} and
  ~\ref{tabupperlimits}, are significantly revised from the 2000-2004
  publications \cite{Boc00,Cro04,Cro04b}:
  \begin{itemize}
  \item Production rates are computed with a code taking
    into account radiative decay, opacity of the lines and a gaussian beam
    with pointing offset, which was not the case in older publications
    \cite{Cro04,Cro04b};
  \item Pumping of the vibrational bands by solar infrared radiation 
    is taken into account for CO, H$_2$CO, CH$_3$OH and CH$_3$CN;
  \item For all other COMs, excepted (CH$_2$OH)$_2$ and CH$_2$OHCHO, molecular
    data \cite{JPL,CDMS} have been revised since the 2000-2004 papers have been
    published, affecting the retrieved abundances by up to a factor of two
    (e.g. C$_2$H$_5$OH \cite{Biv15});
  \item In addition, for CH$_2$CO and c-C$_2$H$_4$O the
    original abundances \cite{Cro04} were off by a factor 2 and 0.1
    respectively, due to a mistake in the calculations;
  \end{itemize}
  Nevertheless, future updates of the molecular abundances are still possible
  as more accurate molecular parameters and photodissociation lifetimes get available.
  A more in-depth analysis of the archive data may also provide somewhat
  different values. Sample spectra
  of recent detections of COMS are presented in Fig.~\ref{figspeccoms}.
  New detections in comets 46P and  C/2013~US$_{10}$ are shown and from
  archive spectra a $4-\sigma$ detection of ethanol in comet C/2013~R1
  is obtained with an abundance comparable to that in comets
  C/2014~Q2 and 46P (Table~\ref{tabcoms}).

\begin{figure}
\centering
\resizebox{\hsize}{!}{\includegraphics[angle=270]{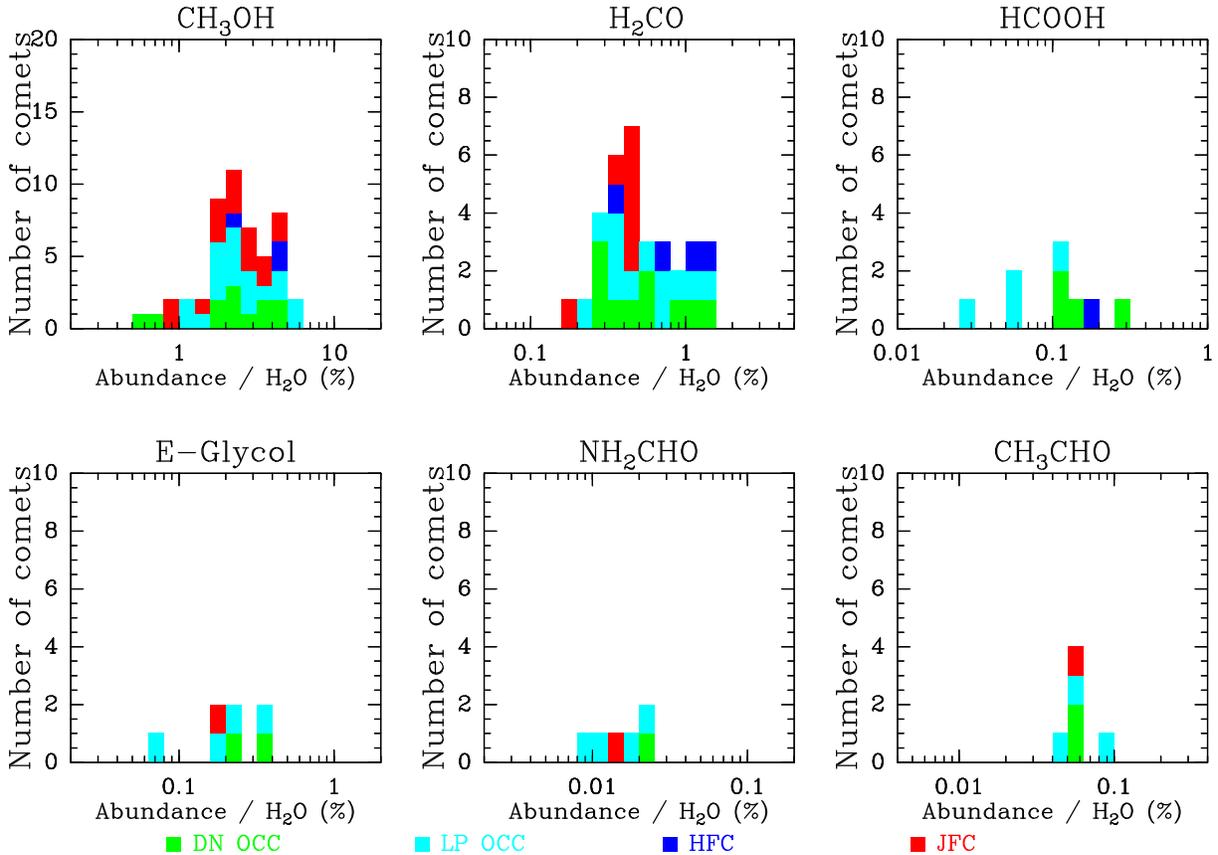}}
\caption{Distribution of the abundances of COMs detected in several comets.
  Comets are color-coded according to their dynamical category: Dynamically
  New Oort Cloud Comets (DN OCC, green), Long-Period OCC (light blue),
  Halley Family Comets (dark blue, originating also from the Oort cloud)
  and Jupiter Family (ecliptic) Comets (red).}
\label{fighistocoms}
\end{figure}

Fig.~\ref{fighistocoms} shows the histogram distribution
of the abundance relative to water of several COMS
detected in more than 6 comets. As the sample of comets
increases, the range of measured abundances relative to water broadens
to reach one order of magnitude variation. Short-period (Jupiter family)
comets (JFC) are poorly sampled due to their limited
intrinsic activity, but do not show significant
differences with Oort cloud comets (OCC) regarding to COMs abundances.
Only the very volatile species CO
seems on average less abundant in JFCs (Table~\ref{tabcoms}).

\begin{table*}
\caption[]{Abundance of organic molecules detected in comets}\label{tabcoms}
\begin{center}
\begin{tabular}{llllll}
\hline\hline
Molecule & Name &\multicolumn{4}{c}{Abundance relative to water in \%} \\  
     & & C/1995~O1   & C/2014~Q2 & in Other & in JFCs \\
     & & (Hale-Bopp) & (Lovejoy) &  OCCs &      \\
\hline
CO$^ a$ & Carbon Monoxide &  $23\pm5$      & $1.8\pm0.2$     & 0.4-22  &  0.3-4    \\ 
H$_2$CO  & Formaldehyde   &  $1.1\pm0.3$   & $0.32\pm0.03$   & 0.24-1.4   &  0.16-0.40 \\
CH$_3$OH & Methanol       &  $2.4\pm0.3$   & $2.4\pm0.1$     & 0.7-6.1    &  0.8-5.0   \\
HCOOH    & Formic Acid    & $0.05\pm0.01$  & $0.028\pm0.003$ & 0.05-0.3   &  $<0.036$ \\
CH$_3$CHO & Acetaldehyde  & $0.08\pm0.01$  & $0.047\pm0.002$ & 0.05-0.06  & $0.06\pm0.01$ \\
(CH$_2$OH)$_2$& Ethylene Glycol & $0.19\pm0.01$  & $0.075\pm0.008$ & 0.24-0.35  & $0.19\pm0.03$ \\
HCOOCH$_3$ & Methyl Formate & $0.065\pm0.009$& $0.080\pm0.011$ & $<0.16$    & $<0.14$  \\
CH$_2$OHCHO & Glycolaldehyde &  $<0.04$       & $0.016\pm0.003$ & $<0.07$    & $<0.04$  \\
C$_2$H$_5$OH & Ethanol &  $<0.20$       & $0.124\pm0.011$ & $0.13\pm0.03$    & $\leq0.11$ \\
NH$_2$CHO    & Formamide & $0.012\pm0.001$& $0.008\pm0.001$ &0.016-0.022 & $0.015\pm0.002$ \\
CH$_3$CN     & Methyl Cyanide & $0.020\pm0.004$& $0.014\pm0.001$ & 0.008-0.054 & 0.017-0.036 \\
\hline
\end{tabular}
\end{center}
Note: $^a$: Including values obtained from infrared surveys\cite{Del16}.
All other abundances are based on radio observations.
For single measurements, the uncertainty on the retrieved abundance is
  provided, taking into account variation with time for the three first molecules.
\end{table*}

\section{Upper limits on undetected species}

Besides the detection of 25 molecules in comets \cite{Biv15}, the radio
observations enabled to derive significant upper limits on the abundance of
other complex molecules\cite{Cro04}. Most of these
molecules have been already detected in the ISM\cite{Bel13,Jor16}. 
Table~\ref{tabupperlimits} provides the most significant upper limits
(or marginal detection).

\begin{table*}
\caption[]{Upper limits on the abundance of Molecules searched for in comets}\label{tabupperlimits}
\begin{center}
\begin{tabular}{llccc}
\hline\hline
Molecule & Name & \multicolumn{3}{c}{Abundance relative to water in \%} \\  
     & & C/1995~O1   & C/2014~Q2  & in Other \\
     & & (Hale-Bopp) & (Lovejoy)  & comets    \\
\hline
CH$_2$CO        & Ketene         & $<0.12$    & $0.0078\pm0.0025$ & $< 0.034$ \\
CH$_3$COCH$_3$  & Acetone        & $$         & $0.011\pm0.003$   & $< 0.04$  \\
CH$_3$OCH$_3$   & Dimethyl Ether & $<0.10$    & $< 0.025$         & $< 0.13$  \\
c-C$_2$H$_4$O   & Ethylene Oxide & $<0.023^a$ & $< 0.006$         & $< 0.029$ \\
CH$_3$NH$_2$    & Methyl Amine   & $$         & $< 0.055$         & $< 0.13$ \\
C$_2$H$_3$CN    & Acrylonitrile  & $$         & $< 0.0027$        & $< 0.013$ \\
C$_2$H$_5$CN    & Ethyl Cyanide  & $<0.01$    & $< 0.0036$        & $< 0.014$ \\
NH$_2$CH$_2$COOH I & Glycine I   & $<0.5$     & $< 0.15$          & $< 0.5$   \\
\hline
\end{tabular}
\end{center}
Note: $^a$: corrected value from previous reference\cite{Cro04} (see text).
\end{table*}

Of interest are the relative abundances of isomers: in the case of
C$_2$H$_4$O, the most abundant isomer in comets is acetaldehyde
(CH$_3$CHO), which has been detected in several comets while
significant upper limits on ethylene oxide (c-C$_2$H$_4$O) have
been obtained. The upper limit on the c-C$_2$H$_4$O/CH$_3$CHO ranges from
0.13 to 0.5 (Tables~\ref{tabcoms},\ref{tabupperlimits}) showing that
acetaldehyde is the main C$_2$H$_4$O isomer in comets, as observed also in the
interstellar medium (ISM) \cite{Bel13}.
Concerning C$_2$H$_6$O, dimethyl ether (CH$_3$OCH$_3$) is also 
not a major isomer, compared to ethanol which is at least five times more
abundant in comet C/2014~Q2 (Lovejoy). The last comparison that can be done
is between methyl formate (HCOOCH$_3$) and glycolaldehyde (CH$_2$OHCHO), the
first one being the most abundant isomer of C$_2$H$_4$O$_2$ in
comets Hale-Bopp and Lovejoy\cite{Biv15} (Table~\ref{tabcoms}).

\section{Discussion}
\begin{figure}
\centering
\resizebox{\hsize}{!}{\includegraphics[angle=270]{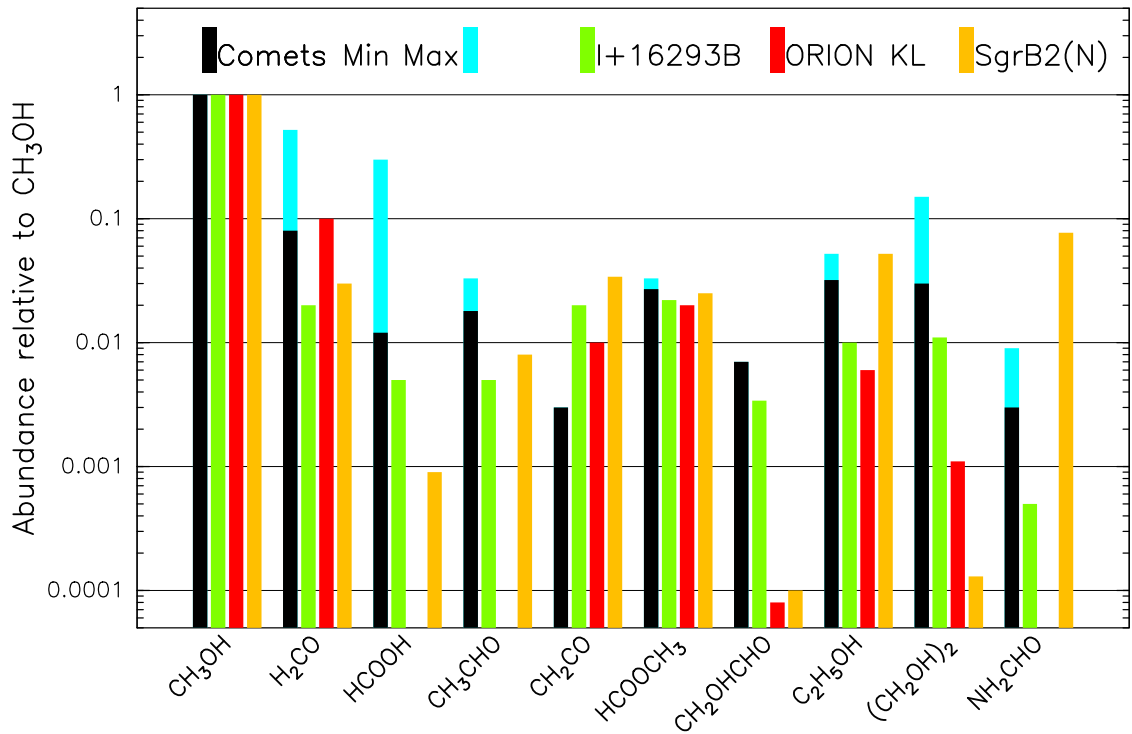}}
\caption{Abundances relative to methanol of organic and complex organic
  molecules in comets and protostars.
  The light blue bar gives the range of values measured in comets with
  the black one setting the minimum value.
  The column density ratios in low-mass protostar IRAS 16293-2422(B)
  are in green\cite{Jor12,Bis08,Jor16}, those for the high-mass protostar
  Orion-KL in red (where available)\cite{Bro15,Bro13} and
  abundances in Sagittarius B2 North source in orange\cite{Bel13}.}
\label{fighistoism}
\end{figure}

Diversity in chemical composition is present in the comet population, but
with few exceptions\cite{Biv18} or in extreme cases (e.g.
observations far from the Sun where the volatility of the molecules can modify
significantly the relative abundances in the coma in comparison to what is
measured within 2 AU from the Sun), relative abundances 
are the same in all comets within one order of magnitude (excepted for the
most volatiles like CO and N$_2$). Abundances of COMs in the ISM
can vary more significantly, depending on the type of sources investigated.
Since comets are remnants of the proto-planetary disk of our solar system,
comparison to the composition of protostars and molecular clouds is instructive.
Fig.~\ref{fighistoism} shows the comparison between the range of
abundances relative to methanol observed in comets and low-mass, high-mass
protostars and SgrB2(N) molecular cloud. Similar abundances are observed and
the scatter between the various ISM sources can often be larger than within
the comet population.
However some molecules like ethylene glycol and acetaldehyde seem more
systematically enriched in comets while ketene, marginally present in comet
C/2014~Q2 (Lovejoy), looks depleted at least in this comet (other upper
limits are compatible with ISM values).

\section{Conclusion}
Since the last thirty years several organic and complex organic molecules
have been detected in comets. The scatter in abundances (relative to water
or methanol) of the COMs is less than one order of magnitude and the abundances
do not seem to depend on the dynamical family (OCC or JFC) of the comet.
Similar abundances are observed within different ISM sources, especially
protostars, sustaining the hypothesis that comets inherited material from
the solar system presolar cloud or formed in the outer protoplanetary disk
in similar physical conditions. However some of the molecules, like e.g.
  glycolaldehyde, have been only observed in one or very few comets
  and further observations of COMs in comets are needed to confirm the
  trends noticed here.

\begin{acknowledgement}
  A large part of this work was based on observations carried out with the
  IRAM 30-m telescope. IRAM is supported by INSU/CNRS (France),
  MPG (Germany) and IGN (Spain).
This research has been supported by the Programme national de 
plan\'etologie de l'Institut des sciences de l'univers (INSU).

\end{acknowledgement}

\newpage


\begin{figure}
\centering
\resizebox{\hsize}{!}{\includegraphics[angle=0]{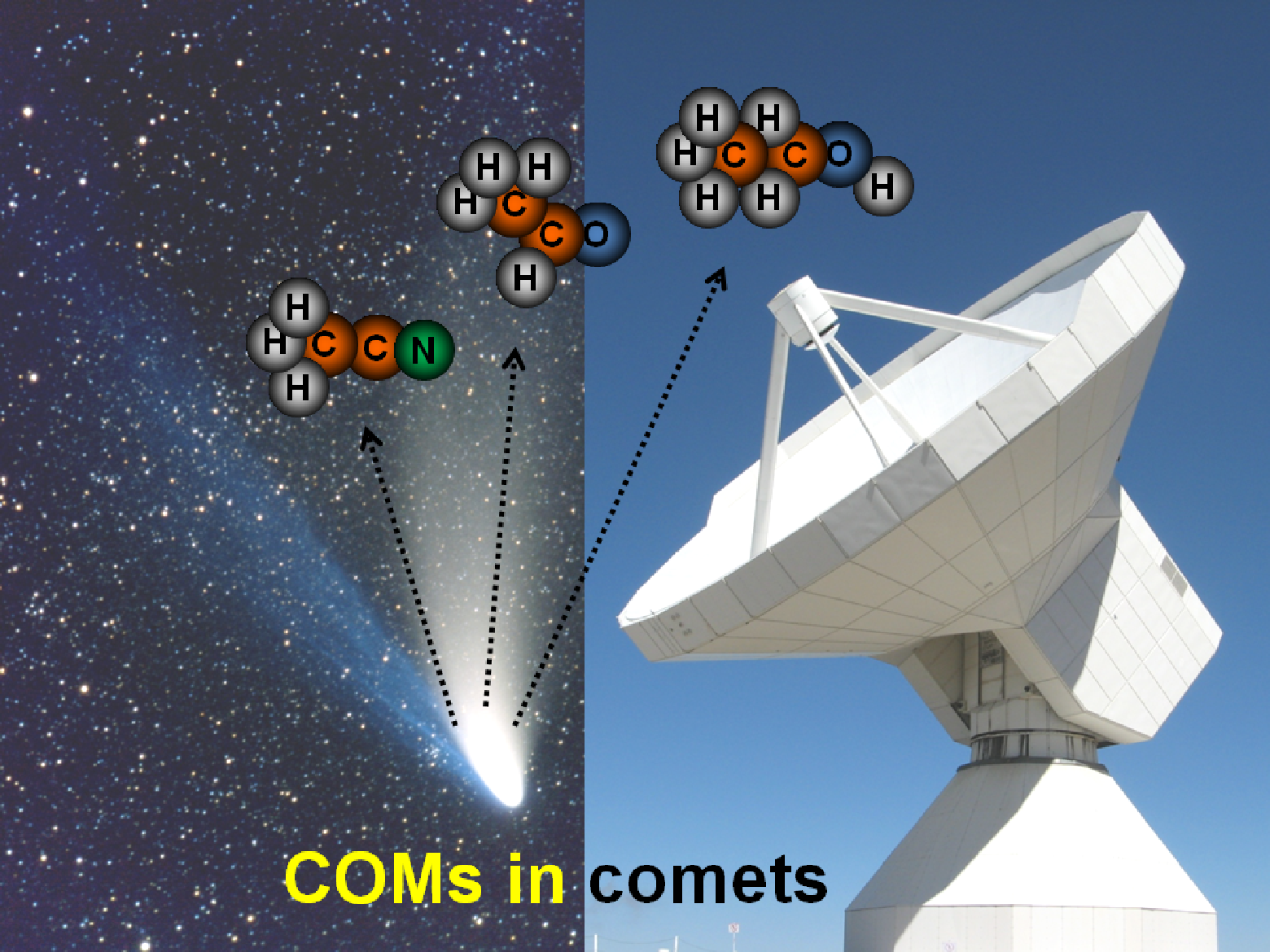}}
\caption{Table Of Content Graphic}
\label{}
\end{figure}

\end{document}